\documentclass[aip,jcp,groupedaddress,superscriptaddress,twocolumn]{revtex4}

\usepackage{amsmath}
\usepackage{graphicx}
\usepackage{dcolumn}
\usepackage{bm}
\usepackage{alltt}%
\usepackage{color,soul}
\usepackage[hidelinks]{hyperref}
\definecolor{orange}{rgb}{0.1,0.99,0.95}
\sethlcolor{orange}

\newcommand{\mX}{\mathcal{X}}

\newcommand{\mM}{\mathcal{M}}

\begin{document}

\title{$\mu$-tempered metadynamics: Artifact independent convergence times for wide hills}

\author{Bradley M. Dickson}
\email{bradley.dickson@vai.org}
\affiliation{Center for Epigenetics, Van Andel Research Institute, Grand Rapids Michigan 49503, USA.}

\date{\today}

\begin{abstract}
Recent analysis of well-tempered metadynamics (WTmetaD) showed that it converges without mollification artifacts in the bias potential. Here we explore how metadynamics heals mollification artifacts, how healing impacts convergence time, and whether alternative temperings may be used to improve efficiency. We introduce ``$\mu$-tempered'' metadynamics as a simple tempering scheme, inspired by a related mollified adaptive biasing potential (mABP), that results in artifact independent convergence of the free energy estimate. We use a toy model to examine the role of artifacts in WTmetaD and solvated alanine dipeptide to compare the well-tempered and $\mu$-tempered frameworks demonstrating fast convergence for hill widths as large as $60^{\circ}$ for $\mu$TmetaD.

\end{abstract}

\maketitle
\section{Introduction}

Metadynamics was proposed in 2002\cite{lp02} as a method for escaping metastabilities, and recently a proof of convergence has emerged\cite{dama2014} for tempered versions, which speaks directly to the form metadynamics takes when it is implemented. A critical feature of that proof, which has driven this report, was that WTmetaD converges without mollification artifacts in the bias potential. The metadynamics literature has not been clear on this point, with discussion of ``blurring''\cite{lrgcm05} and even demonstrations of failed convergence on practical timescales where ``blurring'' is observed\cite{adaptivemeta}. It is an intriguing insight to find that WTmetaD, and the metadynamics family, converges without mollification artifact as it means the hill width does not impact the accuracy of free energy estimation. 

This insight was not predicted in Ref. \onlinecite{pre11}, where we derived a novel bias identical to the motivating expression for well-tempered metadynamics\cite{bbp08}. In Ref. \onlinecite{pre11} we showed that this motivating expression defines a bias potential which converges to an end state that includes mollification artifacts. We also showed how to estimate the exact free energy from the bias potential via deconvolution.

Here, we start out by looking at the disconnect between mABP and WTmetaD. 
The two methods have identical motivating expressions for the bias potential, and agree when the hill width is small, yet diverge from one another when the hill width is large: WTmetaD must heal artifacts in the bias to converge while the mABP bias converges independently of artifacts. Below it is estimated that the timescale for healing mollification artifacts scales exponentially with the hill width for WTmetaD. Motivated to find a tempering schedule that improves upon the expensive healing process, we generalize the tempering of metadynamics into a common framework and propose an alternative schedule. The new scheme, called $\mu$-tempering, produces a free energy estimate that is independent of the bias potential once the hills are small enough to allow the dynamics to be near equilibrium. In other words, artifacts do not need to be healed. We demonstrate that this tempering overcomes the slow convergence of WTmetaD when $\alpha$ is large, allowing use of aggressive filling rates.

\section{Results}
\subsection{Metadynamics from mABP}

The point of this section is to point out the disconnect between mABP and WTmetaD, and to restate a main result of mABP, which was that accurate free energy estimates could be made while employing arbitrary hill widths and one need not spend any time healing artifacts in the bias potential.

We first define some notation and briefly revisit prior results. We write $\bm{x}$ for a single configuration in the $n$-dimensional configuration space $\mX$ of a dynamical system. Assume $N$ collective variables (CVs) are specified and, of course, let these CVs be a good representation of $\mX$. $\Omega$ is the space of CVs and $\bm{\xi}$ is a point in $\Omega$. We define the free energy estimate (up to a constant $\zeta$) 
\begin{equation}
  \label{molDOS}
\begin{split}
 &\zeta e^{-\beta A_{\alpha}(\bm{\xi)}}= \\ 
&Z^{-1}\int_{\mX} \delta_{\alpha}(\bm{\xi}(\bm{x})-\bm{\xi}) \, e^{-\beta V(\bm{x})} \, d\bm{x}, 
\end{split}
\end{equation} where 
\begin{equation}\delta_{\alpha}(\bm{\xi})=\exp\left(-\frac{|\bm{\xi}|^2}{\alpha^2}\right) 
\end{equation} and $\beta=k_BT$. We let $Z$ take the normalization of $\delta_{\alpha}$. 

In reference \onlinecite{pre11} we derived the following biasing potential from an intuitive on-the-fly reweighting scheme 
\begin{equation}\label{apfm}
V_b(\bm{\xi},t) = \beta^{-1} \frac{b}{1-b}\ln[c\,(1-b)\,\displaystyle h_{\alpha}(\bm{\xi},t) +1]
\end{equation} 
where 
\begin{equation}\label{halpha}
h_{\alpha}(\bm{\xi},t) = \displaystyle\int_0^t\delta_{\alpha}(\bm{\xi}(\bm{x}_s)-\bm{\xi})\, ds
\end{equation} This biasing potential was designed to facilitate estimation of $A_{\alpha}$ by enhancing exploration of configuration space. We elaborated that the potential of mean force (PMF) in the biased ensemble is related to the bias potential as $V_b = -b\mu_{\alpha}/(1-b)$ where at long times $\mu_{\alpha} = -\beta^{-1} \ln[ h_{\alpha} ]= -\beta^{-1} \ln[ \delta_{\alpha}\ast h]$. $h$ is a histogram with vanishingly small $\alpha$, and $\ast$ indicates a convolution. Thus the exact PMF is $\mu = -\beta^{-1} \ln[h] = A+V_b$. We showed that the free energy could be estimated from these dynamics as 
\begin{equation}\label{exact}
A(\bm{\xi}) = \mu(\bm{\xi}) + \frac{b}{1-b}\mu_{\alpha}(\bm{\xi})
\end{equation} 
When one can take $\mu_{\alpha}\approx \mu$ the free energy estimate becomes 
\begin{equation}\label{free1}
A(\bm{\xi},t) = -\frac{1}{b}V_b(\bm{\xi},t)
\end{equation} suggesting that $100\times b$ percent of the free energy is cancelled by the bias potential. When $\alpha$ is so large that $\mu_{\alpha}$ cannot replace $\mu$, 
\begin{equation}\label{decon}
A(\bm{\xi},t) = -\beta^{-1}\ln\big[\delta^{-1}_{\alpha}\ast h_{\alpha}(\bm\xi,t) \times h_{\alpha}(\bm\xi,t)^{\frac{b}{1-b}}\big] 
\end{equation} 
where $\delta_{\alpha}^{-1}\ast f$ indicates a deconvolution. The nice thing about this approach is that at equilibrium the free energy estimate simply states $A=A+V_b-V_b$ which is free from artifacts in the bias potential. 

In practice one may collect a histogram on the same domain as $h_{\alpha}$, called $h_{\circ}$, and take the approximation $h_{\circ}\approx \delta_{\alpha}^{-1}\ast h_{\alpha}$ into equation \eqref{decon}.
This histogram may be collected as a standard histogram or it may be collected the same way as $h_{\alpha}$ but with a very small Gaussian width. At equilibrium $h_{\circ}\propto e^{-\beta \mu}$ so again the free energy estimate is free of artifacts caused by the bias and one does not need to wait for artifacts to be healed.

In Ref. \onlinecite{pre11} we also noted that equation \eqref{apfm} maps onto the biasing potential used to motivate well-tempered metadynamics given the following parameter conversions, 
\begin{equation}\label{convert}
\begin{split}
&\omega = \beta^{-1} c\,b \\
&\Delta T = \beta^{-1} b /(1-b)
\end{split}\end{equation} 
Plugging these into equation \eqref{apfm} gives 
\begin{equation}\label{wtmeta}V_b(\bm{\xi},t) = \Delta T\ln \left(\frac{\omega}{\Delta T} h_{\alpha}(\bm\xi,t) +1\right) 
\end{equation} which is equation 2 in the introduction of well-tempered metadynamics, Ref. \onlinecite{bbp08}. 

Plugging the WTmetaD parameters into equation \eqref{free1} gives 
\begin{equation}\label{free2}
A(\bm{\xi},t) = -(1+\frac{\beta^{-1}}{\Delta T})V_b(\bm{\xi},t)
\end{equation} which is exactly the WTmetaD rule for computing free energy from the bias potential. 

The WTmetaD bias potential update is motivated by considering the time derivative
\begin{equation}\label{timed}
\begin{split}
\dot{V}_b(\bm{\xi},t) &= \frac{\omega \delta_{\alpha}(\bm{\xi}(\bm{x}_{t})-\bm{\xi})}
{\left(\frac{\omega}{\Delta T}h_{\alpha}(\bm{\xi},t) +1 \right) }\\
&=
\omega e^{-V_b(\bm{\xi},t)/\Delta T}\delta_{\alpha}(\bm{\xi}(\bm{x}_{t}) - \bm{\xi})
\end{split}\end{equation} This is equation 3 of Ref. \onlinecite{bbp08} and defines the ``hill height'' to be $e^{-V_b(\bm{\xi},t)/\Delta T}$. 

Based on equation \eqref{timed}, 
we can define a bias update with the finite difference 
\begin{equation}\label{timed2}
\dot{V}_b(\bm{\xi},t+dt) = \frac{1}{dt}\left(
V_b(\bm{\xi},t+dt) - V_b(\bm{\xi},t) \right)
\end{equation} to obtain 
\begin{equation}\label{timed4grid}
\begin{split}
& V_b(\bm{\xi},t+dt) = V_b(\bm{\xi},t) + \\
& dt \,\omega e^{-V_b(\bm{\xi},t+dt)/\Delta T}\delta_{\alpha}(\bm{\xi}_{t+dt}-\bm{\xi})
\end{split}
\end{equation} where $\bm\xi_{t+dt}\equiv \bm\xi(\bm x_{t+dt})$ is the center of the most newly deposited hill. This is the update consistent with equation \eqref{wtmeta}, and is simply a numerical solution to equation \eqref{timed}. 

It can be said that this update ``drops'' the Gaussians $\delta_{\alpha}$ along the trajectory, where the Gaussian heights are scaled by $e^{-V_b(\bm{\xi},t+dt)/\Delta T}$. Notice that if the effects of finite $\alpha$ are ignored then it is straight forward to conclude that the limiting equilibrium distribution of $\bm\xi$ generated by equation \eqref{timed4grid} is $e^{-V_b/\Delta T}e^{-\beta (A+V_b)}$. This distribution is uniform on $\Omega$ exactly when equation \eqref{free2} is satisfied. Thus, we concluded in Ref. \onlinecite{pre11} that results pertaining to mABP could be extended to WTmetaD, generating what we called a ``parameter-free'' approach. The idea was that equation \eqref{decon} can correct for $\alpha$-related errors which makes the method exact for any parameter choice provided sufficient sampling. It was known at the time that large hills caused deteriorated convergence of metadynamics.\cite{adaptivemeta}

However, a distinction of methods can be made by considering the {\it implemented} update for WTmetaD 
\begin{equation}\label{timed4}
\begin{split}
&V_b^{WTM}(\bm{\xi},t+dt) = V_b^{WTM}(\bm{\xi},t) + \\
&dt \,\omega e^{-V_b^{WTM}(\bm{\xi}_{t+dt},t)/\Delta T}\delta_{\alpha}(\bm{\xi}_{t+dt}-\bm{\xi})
\end{split} 
\end{equation} 
which is equation 1 of Ref. \onlinecite{dama2014}. 
In Ref. \onlinecite{bbp08} ``s'' is written for $\bm\xi$ and ``s(t)'' is written for $\bm\xi_{t+dt}$. Notice that in Ref. \onlinecite{bbp08} ``s(t)'' does not appear in the hill height. To obtain the update in equation \eqref{timed4} from the update in equation \eqref{timed4grid}, $\bm\xi_{t+dt}$ must be substituted for $\bm\xi$ in the hill height and nowhere else. The consequences of this substitution have never been formally addressed and are not always clearly implied by notation convention in the literature. This slight notational difference actually distinguishes two very different biasing potentials.

WTmetaD can obtained from mABP under the exchange 
\begin{equation}\label{implas}
e^{-V_b(\bm{\xi},t+dt)/\Delta T} \Rightarrow
e^{-V_b(\bm{\xi}_{t+dt},t)/\Delta T}
\end{equation} which redefines the hill height. In reference \onlinecite{pre11} we showed how to shift the point of evaluation in exponential factors like these and, borrowing those results, this exchange remains faithful to mABP and the motivating expression \eqref{timed} when $\alpha$ is small so that mollification artifacts do not enter the bias potential. The differences between these bias potentials only emerges when $\alpha$ is large and is summarized by two observations. First of all, the convergence time of WTmetaD begins to suffer since artifacts must bee healed prior to convergence.\cite{dama2014,adaptivemeta}. Second, mABP becomes increasingly less efficient as $\alpha$ is increased because the mollification artifacts prevent the bias from flattening the landscape. 

The goal of the following sections is to obtain an artifact-independent free energy estimate for metadynamics that is applicable to the {\it implemented} metadynamics. This would enable the use of larger hill widths, providing better sampling enhancement, and give confidence that artifacts are irrelevant for any particular application. The first step though is to get a better idea of the role of the hill width in determining the long-time limit of the bias, as so far the literature demonstrates that the method converges without artifact while at the same time demonstrating that it is impractical to converge exactly in the face of artifacts.

\subsection{Consistency for tempered metadynamics}

It is straightforward to cast WTmetaD in an alternative form by thinking of the update as a time integral and inserting a Dirac delta function. To do this, we partition the time $t$ into $n$ intervals of $dt = t/n$, where $t_1 = 1 dt$, $t_2=2 dt$, $t_n = n dt = t$, and define the weighted histogram 
\begin{equation}\label{disct}\begin{split} &w(\bm\xi,t) = \\
&\lim_{n\rightarrow \infty}\sum_{i=1}^{n} e^{-V_b^{\text{WMT}}(\bm\xi(\bm x_{t_i}),t_i-dt)/\Delta T}\delta(\bm\xi -\bm\xi(\bm x_{t_i})) dt\\
&=\int_{0}^t e^{-V_b^{\text{WMT}}(\bm\xi_s,s)/\Delta T}\delta(\bm\xi -\bm\xi(\bm x_s))ds 
\end{split}\end{equation} 
The WTmetaD bias can now be expressed as 
\begin{equation}\label{RBFaMol}
V_b^{\text{WTM}}(\bm\xi^*,t) = \omega\int_{\Omega}w(\bm\xi,t)\delta_{\alpha}(\bm\xi-\bm\xi^*)d\bm\xi
\end{equation} 
The standard WTmetaD is recovered by changing integration order, integrating $\bm\xi$, and discretizing time as shown in equation \eqref{disct}. The bias is now expressed in terms of a histogram, and this histogram affords some leverage towards understanding the long-time limit of the bias and its dependence on $\alpha$.

With equation \eqref{RBFaMol}, it is possible to make a consistency check by assuming that the bias allows the dynamics to reach equilibrium. Suppose that at time $\tau$ a WTmetaD simulation has completely converged and that simulation times beyond $\tau$ only produce an equilibrium trajectory.

Splitting the histogram as follows 
\begin{equation}\label{eplit}\begin{split}
&w(\bm\xi,t) = w(\bm\xi,\tau)+\\
&\int_{\tau}^t e^{-V_b^{\text{WMT}}(\bm\xi(\bm x_s),s)/\Delta T}\delta(\bm\xi -\bm\xi(\bm x_s))ds \\
&=w(\bm\xi,\tau)+\rho(\bm\xi,t-\tau)
\end{split}\end{equation}
allows us to write the bias as 
\begin{equation}\label{rest}\begin{split}
&V_b^{\text{WTM}}(\bm\xi,t) = V_b^{\text{WTM}}(\bm\xi,\tau) + \\
&\omega\int_{\Omega} \rho(\bm\xi^*,t-\tau)\delta_{\alpha}(\bm\xi^*-\bm\xi) d\bm\xi^* 
\end{split}
\end{equation} 

Since we have assumed equilibrium sampling, we expect that $\rho \propto e^{-V_b^{\text{WTM}}/\Delta T}e^{-\beta \mu}e^{-\max[V_b^{\text{WTM}}]}$. This last factor indicates that the drifting zero-of-energy for the bias is sending the hill heights to zero regardless of robust sampling. Equation \eqref{rest} clearly implies that the system can only come to equilibrium if $\rho\ast \delta_{\alpha}=0$, or $\rho\ast \delta_{\alpha}=const\times e^{-\max[V_b^{\text{WTM}}]}$, leaving the bias potential conservative. For the non-trivial case, $\rho$ must be constant across $\Omega$. From the definition of $\mu$ we see that $\rho$ is constant when $V_b^{\text{WTM}}=-A/(1+\beta^{-1}/\Delta T)$. Thus, WTmetaD produces an exact free energy without any mollification artifacts assuming convergence. 

This conclusion is identical to what we speculated above (while ignoring the finite hill width) and it is identical to the conclusion of Ref. \onlinecite{dama2014}. While we arrive at equation \eqref{rest} via an alternative route, the equation is just a restatement of equation 7 in Ref. \onlinecite{dama2014} and to aid the comparison we have used $\rho$ to indicate the reweighted equilibrium distribution which was written as $e^{-V_b^{\text{WTM}}/\Delta T}\rho_b$ there. ($\rho=e^{-V_b^{\text{WTM}}/\Delta T}\rho_b$) Here, we have assumed convergence and checked what that converged state must be. Dama and co-workers\cite{dama2014} have provided arguments that WTmetaD does converge by further analysing $\dot{V}_b^{\text{WTM}}=\rho\ast \delta_{\alpha}$. 

At this point, knowing that WTmetaD converges without mollification artifact begs the question of how this is achieved in practice, and how this impacts convergence times. 

\subsection{Mollification artifacts and convergence time}
Equation \eqref{RBFaMol} shows that the bias potential is related to a Gaussian ``blurring'' of $w$. If the bias converges exactly to the underlying free energy, then $w$ must take very specific shapes any time $\alpha$ is large compared to the features of the free energy. In particular, a Gaussian filter will reduce the amplitude of features having spatial frequencies that are large compared to $\alpha$. Since WTmetaD converges to the free energy regardless of $\alpha$, some means of compensating for this attenuation of high frequencies must be present.

The long-time limit of WTmetaD gives us
\begin{equation}\label{mini}
\frac{-A(\bm\xi^*)}{\omega(1+\frac{\beta^{-1}}{\Delta T})}
=\int_{\Omega}w(\bm\xi,\infty)\delta_{\alpha}(\bm\xi-\bm\xi^*)d\bm\xi
\end{equation} after combining equation \eqref{RBFaMol} and the relationship between the free energy and bias potential. 
Thus, the histogram that is built during WTmetaD simulations, $w$, must display the high frequency shapes of $A$ with exaggerated amplitude or those shapes will be washed out by the convolution with $\delta_{\alpha}$. 

We can follow this through with an analytic model. While more complex issues may exist, the following example gets at the mechanics of avoiding mollification. 
We take a one dimensional collective variable on $[-\infty:\infty]$, where the target free energy is 
\begin{equation}\label{toy}A(\xi) = \cos(\nu\,\xi)\end{equation} 
The effects of $\delta_{\alpha}$ can be computed exactly, 
\begin{equation} \begin{split}
&\int_{-\infty}^{\infty}\delta_{\alpha}(\xi-\xi^*) A(\xi)d\xi\\
&=\alpha\sqrt{\pi}\exp[-(\alpha\,\nu)^2/4]\cos(\nu \,\xi^*)
\end{split}
\end{equation} showing that the attenuation of frequency ($\nu$) is related to how large $\alpha$ is relative to that frequency. 
Up to an additive constant, the histogram that WTmetaD must generate can be written as a function of $\alpha$
\begin{equation}\label{exactw}
w(\xi,\infty) = 
-\exp[\frac{\alpha^2\nu^2}{4}]\cos(\nu\,\xi)/C(\alpha)
\end{equation} where $C(\alpha)=\alpha\sqrt{\pi}\omega(1+\frac{\beta^{-1}}{\Delta T})$. It is straight forward to check that with this histogram $-V_b^{WTM}(\xi,\infty)(1+\frac{k_BT}{\Delta T})=A$. 

When $\exp[\alpha^2\nu^2] \approx 1$ WTmetaD converges with $w(\xi,t)=-\cos(\nu\xi)/C(\alpha)$, up to some additive constant. As $\alpha$ increases, the number of samples required to converge increases as $\exp[\alpha^2\nu^2]$ for each $\xi$. The histogram $w$ can only grow logarithmically with time due to the tempering, yet the density of samples required to converge increases exponentially with $\alpha$. Thus, the WTmetaD convergence time must be of order $\exp[\alpha^2\nu^2]$ for this model system. This explains why slow convergence is observed for WTmetaD when $\alpha$ is large\cite{adaptivemeta}. (Notice that because of $C(\alpha)$ this analysis also suggests that convergence time increases for small $\alpha$ because $1/C(\alpha)$ becomes large.)

This analysis suggests an imbalance in objectives: The number of samples on $\Omega$ that are required to converge are increasing rapidly with $\alpha$ while the importance of the samples is being diminished by that same accumulation. It becomes more and more costly to heal the artifacts as samples are collected, eventually becoming impractical to observe convergence of the free energy estimate. 

In the next section we consider alternate tempering schedules with the idea of improving the rate of convergence by either increasing the speed at which $w$ evolves or by adjusting the target free energy estimate so that mollification does not scale the convergence time.

\subsection{Generalized tempering}

Consider generalizing the weighted histogram $w$ as
\begin{equation}\label{general}
w(\bm\xi,t) = \int_0^t f(\bm\xi,s)\delta(\bm\xi -\bm\xi(\bm x_s))ds 
\end{equation} where the bias is again given by equation \eqref{RBFaMol}. 
If we take $f(\bm\xi,s) = 1$ we have standard metadynamics. When $f=1$ the bias must converge to $-A$ if $\rho\ast \delta_{\alpha}=const$. Of course, because $f$ is a constant, the metadynamics bias never reaches this state. By always adding fixed height hills, metadynamics ensures that $\rho\ast \delta_{\alpha}$ cannot be a constant across $\Omega$. This point is appreciated as the source of the oscillations in the metadynamics free energy estimate and demonstrates that a consistency check is not a proof. 

If we take $f(\bm\xi,s) = e^{-V_b^{\text{WMT}}(\bm\xi,s)/\Delta T}$ we have WTmetaD. Additionally, we can take $f(\bm\xi,s)=f(s)$ where $f(s)$ is a strictly decaying function of time. One special choice of $f(s)$ has been given in TTmetaD (transition-tempered metadynamics\cite{ttmetaD}). The bias potential generated with the TTmetaD update converges to $-A$ but, unlike metadynamics, leads to asymptotic convergence because the updates are vanishing. TTmetaD does suffer from the problem of artifacts, but will converge eventually.



Any given tempering can be implemented as 
\begin{equation}\label{TM}
\begin{split}
&V_b^{TM}(\bm{\xi},t+dt) = V_b^{TM}(\bm{\xi},t) + \\
&dt \,\omega f(\bm{\xi}_{t+dt},t)\delta_{\alpha}(\bm{\xi}_{t+dt}-\bm{\xi})
\end{split} 
\end{equation}
and the free energy can be estimated as 
\begin{equation}\label{TMest}
A(\bm\xi, t) = \beta^{-1}\ln\Big(f(\bm\xi,t)\Big)-V_b^{TM}(\bm\xi,t)
\end{equation} which follows from our consistency check. We write $TM$ just to mean any ``Tempered Metadynamics.'' One can check that this reduces to the WTmetaD rule upon substitution of the well-temered schedule $f=e^{-V_b^{TM}/\Delta T}$.

Notice that if $f(\bm\xi,t) \propto e^{\beta\mu}$ then equation \eqref{TMest} would capture the spirit of the mABP free energy estimate. Plugging this $f$ into equation \eqref{TMest} gives $A=A+V_b^{TM}-V_b^{TM}$. To achieve this, one may collect the histogram $h_{\circ}$ as defined earlier (also subject to $h_{\circ}(\bm\xi,0)=0$) and use the tempering schedule 
\begin{equation}\label{FM}
f(\bm\xi,t) = \frac{1}{rh_{\circ}(\bm\xi,t)+1} \times \left(\max_{\bm\xi\in \Omega}[rh_{\circ}(\bm\xi,t)+1]\right)^m
\end{equation} which has been designed after equation \eqref{timed}. Based on the fact that at long times $h_{\circ}\propto e^{-\beta \mu}$, at equilibrium equation \eqref{TMest} writes $A=\mu-V_b^{\text{TM}}$ which is exactly what we want. This tempering serves to simultaneously shrink the hills and reweight the free energy estimate. Setting $0\leq m <0.5$ allows some control over the decay rate of the hills while still making sure the hills eventually vanish. Recently Fort and co-workers looked at importance sampling and suggested that this sort of decay rate ({\it i.e.} $t^{m-1}$) may be an improvement for importance sampling and tempered metadynamics\cite{TLself}. However, after numerical experiments below we suggest that $m=0$ is always the best option as taking larger values only delays equilibration of this scheme. 

The parameter $r$ plays the role of $\omega/\Delta T$ in WTmetaD or $c(1-b)$ in mABP. The meaning of $\Delta T$ is lost however, so we simply use $0\leq r$. Indeed, the numerical results below indicate that $r=\omega/\Delta T$ is not an optimal choice in general. The smaller $r$, the slower the decay of the hills and when $r=0$ standard metadynamics is recovered. Since $h_{\circ}$ is defined as 
\[
h_{\circ}(\bm{\xi},t) = \displaystyle\int_0^t\delta_{\circ}(\bm{\xi}(\bm{x}_s)-\bm{\xi})\, ds
\]
$r$ has units of $dt^{-1}$. Ultimately $r$ provides a more fine control over the hill decay than does $m$ because it avoids changing the envelope of the hill decay. 

The only concerns for this scheme are that (1) the hills do not decay too fast, leaving the bias to only inefficiently push trajectories over high barriers, and (2) that the hills decay fast enough to let the dynamics equilibrate on a reasonable timescale. The same two concerns exist when specifying any tempered metadynamics parameter set.

We refer to the tempering schedule of equation \eqref{FM} as ``$\mu$-tempered'' metadynamics or $\mu$TmetaD because it uses the PMF of the biased ensemble ($\mu$) to temper. This scheme bypasses the mechanism for healing artifacts and leans on the shrinking hill heights to guarantee that equilibrium will be met. Once in equilibrium the free energy estimate reduces to an identity $A=A$ (assuming $\Omega$ is robustly sampled), which was designed in analogy to the mABP free energy estimate. 

Next, we use numerical experiments to compare WTmetaD and $\mu$TmetaD.

\subsection{Numerical experiments}

\begin{figure}[h]
\includegraphics[width=\columnwidth]{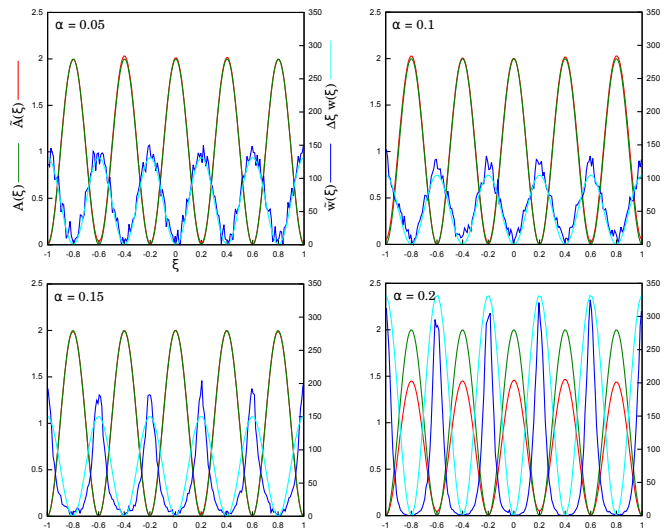}
\caption{Here we demonstrate the impacts of large $\alpha$ and show how $w$ must exaggerate the shape of $A$ to account for the convolution in equation \eqref{RBFaMol}. $\tilde{A}$ and $\tilde{w}$ are the WTmetaD estimates. $\Delta \xi$ is the bin width used in simulations and appears for dimensional consistency. One million MD steps were used.}
\label{EGGS}
\end{figure} 

First we demonstrate the role of $w$ in WTmetaD. We have implemented WTmetaD according to equation \eqref{timed4} so that we may compare WTmetaD simulations to our expectations for the histogram $w$. We used $b=0.8\,,c=0.01$, $k_BT=0.2$ (with no experimentation) and the Langevin integrator from reference \onlinecite{tanda}. The target free energy is given by equation \eqref{toy}. We took $\xi$ on the interval $[-1:1]$ and set $\nu=5\pi$ for this numerical example. We discretized $\Omega$ into 200 bins. 

The histogram $w$ is not normally tracked during WTmetaD simulations but can easily be computed by collecting a standard histogram on $\Omega$ that is reweighted by the hill height scaling. That is, each time a bin on $\Omega$ is visited, the histogram is updated by adding $e^{-V_b^{WTM}/\Delta T}$ to that bin. 

In figure \ref{EGGS}, we plot equation \eqref{exactw} and the histogram collected from a standard WTmetaD simulation implemented as equation \eqref{timed4}, called $\tilde{w}$. As expected, when $\alpha$ increases the high frequency amplitudes of $A$ must be exaggerated in the WTmetaD estimate of $w$. Comparing $\alpha=0.05$ with $\alpha=0.2$, we see that the peaks in $\tilde{w}$ are twice as high at the larger $\alpha$ yet the barriers of the free energy are still under estimated.




\begin{figure*}[ht]
\includegraphics[width=7in]{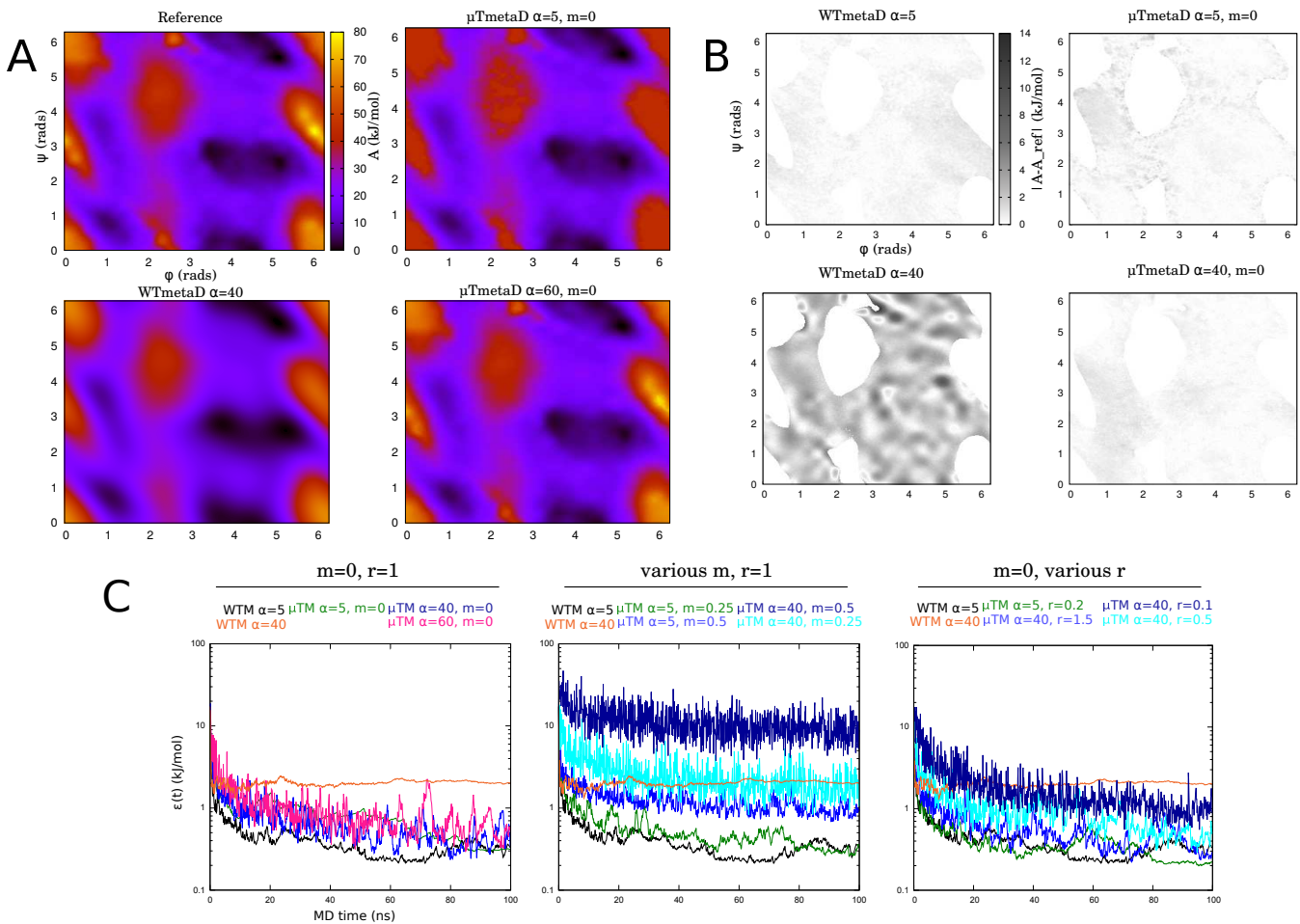}
\caption{(A) The reference free energy is shown along with estimates from the different temperings. (B) The point by point difference between the free energy reference and an estimate for regions contributing to $\epsilon(t)$. (C) The value of $\epsilon(t)$ for the different temperings, and for different $r$ and $m$ values. Rather than $r$ we actually report the unitless value $r\times dt$.
}
\label{freeEs}
\end{figure*}

To look at how $\mu$TmetaD behaves in practice, we considered the case of alanine dipeptide that was used to demonstrate poor convergence for WTmetaD at large values of $\alpha$ in Ref. \onlinecite{adaptivemeta}. In particular it was demonstrated that $\alpha$ can introduce artifacts that are difficult to heal, thus motivating a free energy estimate alternative to equation \eqref{free2} which would also allow the use of more aggressive hill functions. The adaptive Gaussian schemes themselves are unable to improve upon the WTmetaD free energy estimate without abandonment of equation \eqref{free2}. 

Tempered metadynamics, mABP, and a ``SHUS'' scheme based on Ref. \cite{TLself} (see appendix) were implemented in GROMACS-4.5.5\cite{gromacs} and the $\phi - \psi$ dihedral space was broken into a 300-by-300 bin grid. Using CHARMM27 and TIP3P water, Alanine dipeptide was solvated and equilibrated for three nanoseconds in the NPT ensemble. Free energy calculations were done in the NVT ensemble. All codes, inputs, and instructions to reproduce these simulations can be obtained from the ABPenabledGROMACS\cite{megit} github page. 

The biasing parameters were set to $b=0.8$ and $c\,dt=0.1$. ($\omega\,dt=\beta^{-1} c\,dt\,b,\, \text{and}\, \Delta T = \beta^{-1} b /(1-b)$ where $dt$ is the timestep.) Thus, $\omega dt= 0.1992$ and $\Delta T=9.96$ where both are reported in kJ/mol. A reference free energy $A_{\text{ref}}$, was computed from a 100 nanosecond simulation that was biased with a static bias potential. The reference free energy was obtained via equilibrium reweighting. The static bias potential itself was computed from a 10 nanosecond mABP simulation at $\alpha=8$ degrees. The level of convergence was then defined as 
\begin{equation}\begin{split}
\epsilon(t) &= \frac{1}{N}\sum_{i,j} |\hat{A}(\bm\xi_{i,j},t) - A_{\text{ref}}(\bm\xi_{i,j})|\\
&\times \Theta(30-A_{\text{ref}}(\bm\xi_{ij}))\end{split}\end{equation} where the free energy estimate $\hat{A}$ may be derived from $\mu$TmetaD using equation \eqref{TMest} with $h_{\circ}=h_{2^{\circ}}$ or WTmetaD using equation \eqref{free2}. $\Theta$ is the Heaviside step function and the number $N$ is the number of grid points where $A_{\text{ref}} < 30$ kJ/mol. Reference \onlinecite{adaptivemeta} defined convergence in a similar way using nearly the same energy cut-off. Simulations were run with $\alpha=5,40$ degrees for 100 nanoseconds each. The choice $\alpha=40$ was taken from Ref. \onlinecite{adaptivemeta}. All the simulation results are summarized in figure \ref{freeEs}. 

In panel A of figure \ref{freeEs} we show the reference free energy and estimates from different temperings. The estimate from $\mu$TmetaD with $r=1$ and $\alpha=5$ clearly shows that the sampling is inefficient in the high energy regions of the landscape. The ``energy volume'' of each bias update is too small. In this limit the bias updates do more to diminish the hill heights than they do to enhance sampling. The analogous case is setting $\Delta T$ too low in a WTmetaD simulation. In panel C we show that with $r=0.2$ the convergence of $\mu$-TmetaD matches that of WTmetaD when $\alpha=5$. The smaller $r$ allows the hills to decay more slowly but maintains the $1/t$ decay. 

Free energy estimates from WTmetaD with $\alpha=40$ and $\mu$TmetaD with $\alpha=60$ and $r=1$ are also shown in panel A of figure \ref{freeEs}. The smoothing effects of $\alpha$ are visible by eye in the WTmetaD estimate, yet the $\mu$TmetaD estimate is as sharp as the reference even though this simulation used a value of $\alpha$ that is significantly larger. Also notice the improved sampling of high energy regions for the larger $\alpha$, reflecting the larger filling-rate that motivates using wider hills.

In panel B of figure \ref{freeEs} we show the point by point difference in free energy that is used to evaluate $\epsilon(t)$. When $\alpha$ is small, WTmetaD produces a sharp free energy estimate with little error. When $\alpha=40$, WTmetaD produces a far less crisp estimate resulting in point by point errors that can be significant (i.e., up to 5 or 6 $k_BT$). The point by point error in the $\mu$TmetaD result at $\alpha=40$ is, in contrast, still as small as it was when $\alpha=5$. This demonstrates that the free energy estimate is free from $\alpha$-related artifacts, where all simulations have evolved for the same duration (100ns).

Finally, in panel C of figure \ref{freeEs} we show plots of $\epsilon(t)$ for the different temperings and different parameters. Keep in mind that the values of $\epsilon$ are smaller than those shown in panel B simply because $\epsilon$ is an average. What is clear from panel C is that $\mu$TmetaD is converging rapidly for $\alpha=5,\,40,\,60$ (with $m=0$, $r=1$) while WTmetaD is reaching a plateau for larger $\alpha$. This plateau is consistent with the visible differences shown in panel A and the point by point error in panel B, as well as with the previous reports of unobserved convergence at $\alpha=40$. Cleary, $\mu$TmetaD displays artifact independent convergence.

Panel C of figure \ref{freeEs} also shows the behavior of $\mu$TmetaD for $m=0.25,\, 0.5$ when $r=1$. We see that increasing $m$ can impede convergence no matter the size of $\alpha$. The $\mu$TmetaD free energy estimate is only meaningful at equilibrium, where equation \eqref{TMest} can be interpreted as $A+V_b^{TM}-V_b^{TM}$. Taking $m>0$ delays equilibration and clearly has negative impact on convergence time. We observed the same qualitative behavior with the scheme in Appendix equation \eqref{altTM} (results not shown, but the scheme is available in the github\cite{megit} repository). Our summary is that taking decay rates that go as $t^{m-1}$ does not improve convergence and instead only slows equilibration leading to higher levels of noise in the free energy estimate. The source of error in the $\mu$TmetaD free energy estimates stems from uncertainty in the depths of the basins and not in the resolution of the estimate. Since the hills decay very slowly, the noisy behavior of standard metadynamics is partly recovered.

Lastly, panel C shows that $r$ can be used to change the hill decay without the drastic delay of equilibration that is caused by changing $m$. The errors that remain in the free energy when $\alpha=40$, $m=0$, and $r=0.1$ stem from incorrect determination of relative depths of the free energy minima. The free energy estimate itself is as crisp as what is shown in panel A when $\alpha=60$. The tempering is much more robust to adjustments in $r$ than it is to $m$, suggesting $m=0$ is a best choice. Since $r$ simply needs to be a function of time or a constant, a wide range of variations can be imagined and we do not try to find optimal solutions here.


\section{Conclusion}
The difference between mABP and WTmetaD traces back to a single approximation (equation \eqref{implas}) which causes the methods to diverge from one another when $\alpha$ is large. The two main consequences are that the results of mABP do not apply to metadynamics and that metadynamics must heal all artifacts in the bias potential before the free energy can be accurately estimated.

Because the hills of WTmetaD are tempered and shrinking, the process of healing artifacts extends convergence times considerably. The artifacts themselves require more samples accumulated to $w$ and all tempering schedules decrease the weight of each MD timestep when it contributes to $w$. In analogy with mABP, a tempering scheme that does not need to heal artifacts to reach an accurate free energy estimate was proposed, called $\mu$-tempered metadynamics. This scheme was compared to WTmetaD in the common benchmark system of solvated alanine dipeptide. 

The $\mu$TmetaD combines the spirit of mABP with the tempering of metadynamics to achive a method that may represent the best of both worlds: Convergence times that do not scale with $\alpha$, and full exploitation of wide hills. In mABP, configuration space exploration begins to suffer in the limit of large $\alpha$.\cite{pre11} In traditional metadynamics schemes configuration space exploration is improved by large $\alpha$ but the free energy estimate (and convergence) suffers. $\mu$TmetaD allows wide hills to be used for the best sampling efficiencies and produces artifact independent convergence times.

\section{Acknowledgements}
All computations were performed on the Van Andel Research Institute compute cluster.

\appendix

\section{Pseudo codes}\label{pseudo}
Here we will write simple pseudo-code for each adaptive biasing scheme. Each scheme stores information on a grid, which is used to construct the biasing potential and biasing force. To stress the similarity of these methods we define $\mM$ and $\mM'$ as two arrays stored in memory. We store information for the bias potential in $\mM$ and information for the biasing force in $\mM'$.

\noindent {\bf WTmetaD and $\mu$-TmetaD}\newline
Suppose it is time to deposit a new hill. The new hill has center $\bm \xi^* = \bm \xi (\bm x_t)$, and the hill height scale, $s$, of this hill is 
\[ f(\bm\xi^*,t) = \omega \,\exp[ -\mM(\bm \xi^*,t)/\Delta T ] \]
The arrays $\mM$ and $\mM'$ are updated at all grid points $\bm \xi$ as
\begin{equation}\begin{split} 
&\mM(\bm \xi,t+dt) = \mM(\bm \xi,t)+f\,\delta_{\alpha}(\bm \xi^*-\bm\xi) \\
&\mM'(\bm \xi,t+dt) = \mM'(\bm \xi,t)+f\,\delta_{\alpha}'(\bm \xi^*-\bm\xi)
\end{split}\end{equation}
where 
\[\delta_{\alpha}'(\bm \xi^*-\bm\xi)
=\frac{2}{\alpha^2}(\bm\xi^*-\bm\xi)\delta_{\alpha}(\bm \xi^*-\bm\xi) \]
The biasing force at $\bm\xi$ is given by $-\mM'(\bm\xi,t)$. The bias potential is $\mM$. For periodic collective variables, the minimum image convention must be observed. We take the traditional factor of $dt$ into $\omega$ (and the mABP parameter $c$) here.

$\mu$TmetaD is the same as WTmetaD except that 
\[ f(\bm\xi^*,t) = \frac{\omega}{rh_{\circ}(\bm\xi^*,t)+1} \left(\max_{\bm\xi\in \Omega}[rh_{\circ}(\bm\xi,t)+1]\right)^m\]
where $h_{\circ}$ is a histogram as outlined above. 

\noindent {\bf mABP}\newline
Again let $\bm \xi^* = \bm \xi (\bm x_t)$ denote the position of the trajectory in this update. 
The arrays stored in memory are updated at all grid points $\bm \xi$ as
\begin{equation}\begin{split} 
&\mM(\bm \xi,\tau+dt) = \mM(\bm \xi,t)+\delta_{\alpha}(\bm \xi^*-\bm\xi) \\
&\mM'(\bm \xi,\tau+dt) = \mM'(\bm \xi,t)+\delta_{\alpha}'(\bm \xi^*-\bm\xi)
\end{split}\end{equation}
The biasing force at $\bm\xi$ is given by \[\frac{-c\,b\,\beta^{-1} \mM'(\bm\xi,\tau)}{c(1-b)\mM(\bm\xi,t)+1}\] and the bias potential is $\beta^{-1}\frac{b}{1-b}\ln[c(1-b)\mM+1]$ but only the biasing force is evaluated during simulations. 

Ideally $\delta_{\alpha}$ is pre-computed at the beginning of simulation and stored for look-up. All simulations in this work used pre-computed $\delta_{\alpha}$.

\section{mABP as a tempering schedule}

For completeness we note that frequency attenuation may be introduced into the bias potential with the following definition for $f$, 
\begin{equation}\label{unusual} w(\bm\xi,\bm\xi^*,t) = \int_0^t e^{-V_b(\bm\xi^*,s)/\Delta T}\delta(\bm\xi -\bm\xi(\bm x_s))ds \end{equation}
Plugging into equation \eqref{RBFaMol}, the bias becomes
\begin{equation}\label{mABPerk}\begin{split}
V_b(\bm\xi^*,t) 
&=\omega \int_0^t\frac{\delta_{\alpha}(\bm\xi(\bm x_s)-\bm\xi^*)}{\frac{\omega}{\Delta T}\int_0^s\delta_{\alpha}(\bm\xi(\bm x_{\tau})-\bm\xi^*)d\tau+1}ds\\
&=\Delta T\ln \left(\frac{\omega}{\Delta T} h_{\alpha}(\bm\xi^*,t) +1\right) 
\end{split}
\end{equation} which recovers mABP. In passing from the first line to the second line, we solved the differential equation given in equation \eqref{timed}. The bias potential is now a log-histogram so updates to the bias potential decay as $1/t$ even though all observations of $\bm\xi$ contribute equally to $h_{\alpha}$. This $1/t$ decay rate motivated the exponential temperings used in WTmetaD and TTmetaD\cite{ttmetaD}, but requires no tempering or reweighting when the bias is defined directly as a log-histogram. 

Because the solution to the differential equation in equation \eqref{timed} is unique given a boundary condition $V_{b}(\Omega,0)=0$, one may conclude that the only scheme consistent with equation \eqref{timed} is mABP. The WTmetaD update in equation \eqref{timed4} is not strictly a numerical solution to equation \eqref{timed}. This observation does not change the fact that WTmetaD is convergent, and rather argues that equation \eqref{timed} is not useful for describing WTmetaD (or other metadynamics). 

\section{A SHUS-like scheme}
Taking advantage of the flexibility in equation \eqref{general}, we proposed a tempering inspired by ``Algorithm 3'' in reference \onlinecite{TLself}. 
In particular, 
\begin{equation}\label{altTM}
f(\bm\xi,t)=\frac{1}{\ln[V_b^{TM}(\bm{\xi},t)+1]^{\frac{\gamma}{1-\gamma}}+1}
\end{equation}
although this is not identical to ``Algorithm 3'' in Ref. \onlinecite{TLself}. Note that the parameter value inputs for the code on the github page\cite{megit} is the value of $\gamma/(1-\gamma)$.


 
\bibliography{bibs.bib}

\begin{thebibliography}{11}
\expandafter\ifx\csname natexlab\endcsname\relax\def\natexlab#1{#1}\fi
\expandafter\ifx\csname bibnamefont\endcsname\relax
  \def\bibnamefont#1{#1}\fi
\expandafter\ifx\csname bibfnamefont\endcsname\relax
  \def\bibfnamefont#1{#1}\fi
\expandafter\ifx\csname citenamefont\endcsname\relax
  \def\citenamefont#1{#1}\fi
\expandafter\ifx\csname url\endcsname\relax
  \def\url#1{\texttt{#1}}\fi
\expandafter\ifx\csname urlprefix\endcsname\relax\def\urlprefix{URL }\fi
\providecommand{\bibinfo}[2]{#2}
\providecommand{\eprint}[2][]{\url{#2}}

\bibitem[{\citenamefont{Dickson}(2011)}]{pre11}
\bibinfo{author}{\bibfnamefont{B.~M.} \bibnamefont{Dickson}},
  \bibinfo{journal}{Phys. Rev. E} \textbf{\bibinfo{volume}{84}},
  \bibinfo{pages}{037701} (\bibinfo{year}{2011}).

\bibitem[{\citenamefont{Barducci et~al.}(2008)\citenamefont{Barducci, Bussi,
  and Parrinello}}]{bbp08}
\bibinfo{author}{\bibfnamefont{A.}~\bibnamefont{Barducci}},
  \bibinfo{author}{\bibfnamefont{G.}~\bibnamefont{Bussi}}, \bibnamefont{and}
  \bibinfo{author}{\bibfnamefont{M.}~\bibnamefont{Parrinello}},
  \bibinfo{journal}{Phys. Rev. Lett.} \textbf{\bibinfo{volume}{100}},
  \bibinfo{pages}{020603} (\bibinfo{year}{2008}).

\bibitem[{\citenamefont{Laio and Parrinello}(2002)}]{lp02}
\bibinfo{author}{\bibfnamefont{A.}~\bibnamefont{Laio}} \bibnamefont{and}
  \bibinfo{author}{\bibfnamefont{M.}~\bibnamefont{Parrinello}},
  \bibinfo{journal}{PNAS} \textbf{\bibinfo{volume}{99}}, \bibinfo{pages}{12562}
  (\bibinfo{year}{2002}).

\bibitem[{\citenamefont{Dama et~al.}(2014{\natexlab{a}})\citenamefont{Dama,
  Parrinello, and Voth}}]{dama2014}
\bibinfo{author}{\bibfnamefont{J.~F.} \bibnamefont{Dama}},
  \bibinfo{author}{\bibfnamefont{M.}~\bibnamefont{Parrinello}},
  \bibnamefont{and} \bibinfo{author}{\bibfnamefont{G.~A.} \bibnamefont{Voth}},
  \bibinfo{journal}{Physical Review Letters} \textbf{\bibinfo{volume}{112}},
  \bibinfo{pages}{240602} (\bibinfo{year}{2014}{\natexlab{a}}).

\bibitem[{\citenamefont{Laio et~al.}(2005)\citenamefont{Laio, Rodriguez-Fortea,
  Gervasio, Ceccarelli, and Parrinello}}]{lrgcm05}
\bibinfo{author}{\bibfnamefont{A.}~\bibnamefont{Laio}},
  \bibinfo{author}{\bibfnamefont{A.}~\bibnamefont{Rodriguez-Fortea}},
  \bibinfo{author}{\bibfnamefont{F.~L.} \bibnamefont{Gervasio}},
  \bibinfo{author}{\bibfnamefont{M.}~\bibnamefont{Ceccarelli}},
  \bibnamefont{and}
  \bibinfo{author}{\bibfnamefont{M.}~\bibnamefont{Parrinello}},
  \bibinfo{journal}{J. Phys. Chem. B} \textbf{\bibinfo{volume}{109}},
  \bibinfo{pages}{6714} (\bibinfo{year}{2005}).

\bibitem[{\citenamefont{Branduardi et~al.}(2012)\citenamefont{Branduardi,
  Bussi, and Parrinello}}]{adaptivemeta}
\bibinfo{author}{\bibfnamefont{D.}~\bibnamefont{Branduardi}},
  \bibinfo{author}{\bibfnamefont{G.}~\bibnamefont{Bussi}}, \bibnamefont{and}
  \bibinfo{author}{\bibfnamefont{M.}~\bibnamefont{Parrinello}},
  \bibinfo{journal}{Journal of Chemical Theory and Computation}
  \textbf{\bibinfo{volume}{8}}, \bibinfo{pages}{2247} (\bibinfo{year}{2012}).

\bibitem[{\citenamefont{Dama et~al.}(2014{\natexlab{b}})\citenamefont{Dama,
  Rotskoff, Parrinello, and Voth}}]{ttmetaD}
\bibinfo{author}{\bibfnamefont{J.~F.} \bibnamefont{Dama}},
  \bibinfo{author}{\bibfnamefont{G.}~\bibnamefont{Rotskoff}},
  \bibinfo{author}{\bibfnamefont{M.}~\bibnamefont{Parrinello}},
  \bibnamefont{and} \bibinfo{author}{\bibfnamefont{G.~A.} \bibnamefont{Voth}},
  \bibinfo{journal}{Journal of Chemical Theory and Computation}
  (\bibinfo{year}{2014}{\natexlab{b}}).

\bibitem[{\citenamefont{Fort et~al.}(2014)\citenamefont{Fort, Jourdain,
  Lelievre, and Stoltz}}]{TLself}
\bibinfo{author}{\bibfnamefont{G.}~\bibnamefont{Fort}},
  \bibinfo{author}{\bibfnamefont{B.}~\bibnamefont{Jourdain}},
  \bibinfo{author}{\bibfnamefont{T.}~\bibnamefont{Lelievre}}, \bibnamefont{and}
  \bibinfo{author}{\bibfnamefont{G.}~\bibnamefont{Stoltz}},
  \bibinfo{journal}{arXiv preprint arXiv:1410.2109}  (\bibinfo{year}{2014}).

\bibitem[{\citenamefont{Allen and Tildesley}(1989)}]{tanda}
\bibinfo{author}{\bibfnamefont{M.}~\bibnamefont{Allen}} \bibnamefont{and}
  \bibinfo{author}{\bibfnamefont{D.}~\bibnamefont{Tildesley}},
  \emph{\bibinfo{title}{Computer simulation of liquids}}
  (\bibinfo{publisher}{New York: Oxford}, \bibinfo{year}{1989}).

\bibitem[{\citenamefont{Hess et~al.}(2008)\citenamefont{Hess, Kutzner, van~der
  Spoel, and Lindahl}}]{gromacs}
\bibinfo{author}{\bibfnamefont{B.}~\bibnamefont{Hess}},
  \bibinfo{author}{\bibfnamefont{C.}~\bibnamefont{Kutzner}},
  \bibinfo{author}{\bibfnamefont{D.}~\bibnamefont{van~der Spoel}},
  \bibnamefont{and} \bibinfo{author}{\bibfnamefont{E.}~\bibnamefont{Lindahl}},
  \bibinfo{journal}{J. Chem. Theory Comput.} \textbf{\bibinfo{volume}{4}},
  \bibinfo{pages}{435} (\bibinfo{year}{2008}).

\bibitem[{\citenamefont{Dickson}(2015)}]{megit}
\bibinfo{author}{\bibfnamefont{B.}~\bibnamefont{Dickson}},
  \bibinfo{journal}{https://github.com/ BradleyDickson/ABPenabledGROMACS}
  (\bibinfo{year}{2015}), \bibinfo{note}{accessed: 2015-06-13}.

\end{thebibliography}
\newpage


\end{document}